# Electronic conductivity and structural distortion at the interface between insulators SrTiO$_3$ and LaAlO$_3$

**J.-L. Maurice** [*,1], **C. Carretero**[1], **M.-J. Casanove**[2], **K. Bouzehouane**[1], **S. Guyard**[1], **É. Larquet**[3], and **J.-P. Contour**[1]

[1] Unité Mixte de Physique CNRS/Thales, Route départementale 128, 91767 Palaiseau cedex, France
[2] CEMES CNRS, 29 rue Jeanne Marvig, BP 4347, 31055 Toulouse cedex, France
[3] IMPMC, Université P. & M. Curie, Campus Boucicaut, 75015 Paris, France

When insulator LaAlO$_3$ is grown by epitaxy onto a TiO$_2$-terminated {100} surface of insulator SrTiO$_3$, the system obtained has a metallic character. This phenomenon has been associated with an electrostatic frustration at the interface, as {100} surfaces of SrTiO$_3$ are neutral while those of LaAlO$_3$ are polar, but its microscopic mechanism is not quite understood. Here, we present a structural characterisation of this interface by aberration-corrected transmission electron microscopy. The unit cells at the interface appear elongated: we discuss this distortion in terms of electrostatic charge and extra carriers at the interface.

## 1 Introduction

Perovskite oxides are the object of an increased attention due to their versatile properties that could lead to novel applications in microelectronics [1] and spintronics [2]. The generic perovskite oxide unit cell is a simple cube of formula ABO$_3$ where A and B are respectively a large and a small cation. In <100> directions, perovskites are stacks of {200} planes of alternate compositions AO and BO$_2$. There are thus two possibilities of stacking sequence at a {100} interface, which may have dramatically different properties depending on the valency of the different cations in presence [3]. At the interface between the insulators LaAlO$_3$ (LAO) and SrTiO$_3$ (STO) for instance, the -LaO-TiO$_2$- and -AlO$_2$-SrO- sequences should present opposite electronic properties, the former donating, and the latter accepting, electrons [4,5]. Ohtomo and Hwang indeed detected high-mobility electrons in samples made of thin films of LAO grown onto TiO$_2$-terminated STO substrates (-LaO-TiO$_2$- sequence) while they found samples of LAO grown on SrO-terminated STO (-AlO$_2$-SrO- sequence) insulating [5]. In a most recent paper, Nakagawa, Hwang, and Muller presented annular dark field images and electron energy loss spectra recorded with a scanning transmission electron microscope on these interfaces, which indicated that the metallic interface included mixed valency Ti ions, while the insulating one contained oxygen vacancies [6]. We have grown LAO onto TiO$_2$-terminated STO in conditions similar to those used by Ohtomo and Hwang [5]. The aim of this paper is to present our results on interface characterisation at the atomic level, using spherical-aberration ($C_s$) corrected high-resolution transmission electron microscopy (HRTEM). Several of our micrographs exhibit an elongation of the TiO$_6$ octahedra at the interface, an observation which is confirmed by high-angle annular dark field scanning transmission electron microscopy (STEM-HAADF). We discuss this distortion in the images both as the result of an intrinsic distortion of the octahedra in the material, and as the effect of localised charge at the interface.

[*] Corresponding author: e-mail: jean-luc.maurice@thalesgroup.com

## 2 Experimental

We have prepared the -LaO-TiO$_2$- interface by growing LAO onto TiO$_2$-terminated (001) STO using pulsed laser deposition [7,8]. The epitaxy was performed at a temperature of about 750°C and an oxygen pressure of 10$^{-4}$ Pa maintained during cooling; it was controlled *in situ* by reflection high-energy electron diffraction. The thickness of the LAO films was varied from 7 to 22 nm. We used standard photolithography to design a contact pattern on the samples. We compared in each case contacts directly taken on the LAO surface with contacts taken after ion etching a few nm above and below the interface (see Fig. 1). The contacts were made of a stack of Al and Au sputter-deposited films. We obtained *I(V)* curves with a standard four-probe equipment.

We then observed the interface at the atomic level by high-resolution transmission electron microscopy (HRTEM). In conventional HRTEM, the best atomic contrast is obtained at non-zero defocus. At heterointerfaces, such conditions produce Fresnel fringes which may modify the apparent position of atomic columns in the image. Moreover, three-fold astigmatism, which is hardly avoidable with conventional equipment, introduces additional uncertainty [9]. In order to limit such artefacts due to defocus and astigmatism, we used a Tecnai G$^2$ F20 S-Twin equipped with a $C_s$ corrector and automatic fine-tuning of astigmatism. Moreover, we also used a Jeol JEM 2100 equipped with a scanning stage (STEM) and a high angle annular dark field (HAADF) detector. The intensity in STEM-HAADF images depends only weakly on imaging conditions – contrary to HRTEM – so we could use such images as additional proofs that the distortions measured were no artefacts due to the imaging process.

The samples were thinned mechanically using the tripod technique down to a few µm, and then ion milled (Ar, 2 keV, 6°). This process unfortunately delivered surfaces covered with amorphous mater, especially on LAO, which considerably limited the spatial resolution attainable. We prepared cross sections perpendicular to <100> and <110> zones. The 'A' and 'B' cations of the perovskite structure form separate atomic columns in either orientation: A and BO in <100> zone, AO and B in <110>. Oxygen being coupled with a different cation in either case, the respective projected potentials differ not only by geometry, but also by intensity. In fact, the potentials of Sr and TiO columns are about the same for 200-kV electrons, while they differ for SrO and Ti, so that {110} images of STO potentially contain more information. Our {110} samples were however less robust than our {100}, so that their HRTEM images were finally not as significant. In the following, we present an HRTEM image of a {100} sample and an HAADF image of a {110} cross section. We did not attempt to localise pure oxygen columns, which needs very specific sample preparation and imaging conditions [10].

We used a two-fold approach to measure the distortions at the interface. (1) In order to get a statistically relevant result, we used a Fourier analysis of the images (HRTEM and HAADF), in which the local value of the 002 spatial frequency was mapped across the (001) interface [11] (see Fig. 2). Due to the finite size of the mask in this technique, the definition of the maps obtained, and the measured distortions, were spread over two to three lattice parameters (~ 1 nm). (2) So we also analyzed the 002 lattice spacing in real space to get a better idea of the amplitude and localization of the effect, with conversely a larger uncertainty. In all cases, experimental images were part of defocus series and the actual defocus and thickness were determined by comparing with simulations. We used the EMS suite [12] for these calculations.

## 3 Results

X-ray diffraction in the Bragg-Brentano geometry indicated that the lattice parameter of LAO in the growth direction was 0.378 nm, which corresponds to a 0.3 % decrease compared to the bulk parameter. Electron diffraction showed that growth was pseudomorphic in all samples observed, LAO adopting STO's in-plane parameter. Cubic STO has a bulk lattice parameter $a_{STO}$ = 0.3905 nm and the pseudo-cubic unit cell of rhombohedral LAO has a parameter $a_{LAO}$ = 0.3792 nm, so that $(a_{LAO} - a_{STO})/a_{STO}$ = -2.9%. Applying isotropic elasticity theory, the in-plane and out-of-plane strains we have measured give the rather small Poisson ratio of 0.05 for LAO. A possibility is that the equilibrium lattice parameter of our LAO films was larger than that of standard material, which could indicate the presence of oxygen vacancies. The films appeared disoriented by a small fraction of a degree with respect to the substrate, which made it impossible to have both film and substrate in exact zone orientation; it might be related to a systematic bending of the thin foil at the interface. No extended defects could be visualised, neither in HRTEM images nor in low-magnification, large-area pictures. This shows that no plastic relaxation occurred, on the one hand, and that the disorientation, possibly associated with the rhombohedral distortion of LAO (~ 0.06°), did not occasion the presence of dislocations. This is consistent with what we have observed previously in the case of manganites on STO, where the rhombohedral distortion (0.25°, in this case) could be partially recovered with no apparent defects within the first few nm above the interface [13].

Figure 1 shows current-voltage $I(V)$ curves recorded at an interface between a 20-nm thick LAO film and STO: depending on the depth at which the contact is taken, the system appears insulating (metallization separated from the interface by some LAO thickness), or conducting (metallization touching the interface). This conductivity was observed in all the samples analysed and it decreased as temperature increased.

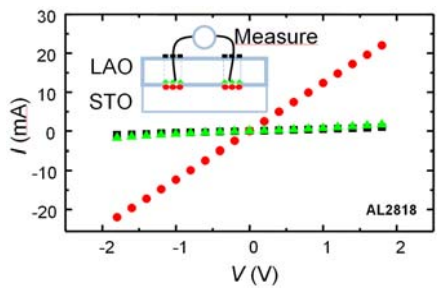

**Fig. 1** (Colour online). Current-voltage curves recorded at 300 K on a LAO/STO sample, with contacts taken at different depth as shown by the schematic in inset. Black squares correspond to contacts taken directly onto the LAO surface, green triangles, to contacts taken into the LAO film close to interface, after ion etching, red circles to contacts taken into STO just below interface. The four points were not aligned so that the $V/I$ ratio corresponds to no meaningful resistance.

Figure 2 summarises the analytical process we applied to HRTEM micrographs and indicates the magnitude and localisation of the distortion at the interface. The variations of the 002 period (red dashed curve in Fig. 2c) show unambiguously that the unit cell pattern undergoes a dilation at the interface.

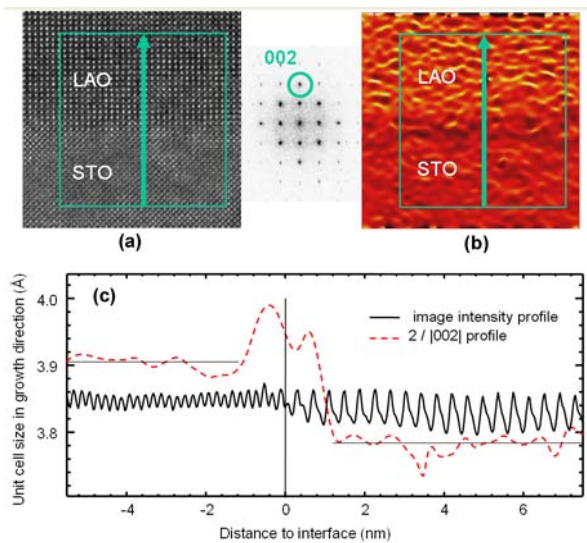

**Fig. 2** (Colour online). The interface in {100} cross section. (a) HRTEM image recorded at 200 kV, $C_s$ = 10 μm, thickness 16 nm, defocus -18 nm. (b) Map of the modulus of the 002 spatial frequency. Scale is given by the arrows along which the intensity profiles displayed in (c) were recorded.

(c) Intensity profiles of (a) and (b), averaged parallel to interface as indicated by the green rectangles in (a) and (b). Twice the inverse of the |002| value has been plotted instead of the direct intensity of (b), in order to better evidence the dilation of unit cell pattern. The horizontal lines give the average values of the parameter in STO (0.3905 nm) and LAO (0.3784 nm); the vertical scale refers only to the 2/|002| profile, the scale of the image intensity is arbitrary.

In the HRTEM image in Fig. 2a, where essentially AlO atomic columns are visible in LAO, the analysis of the intensity profile in real space (solid curve in Fig. 2c) indicates that the dilation is localized on the interface unit cells. Its relative amplitude is +4.9 % while the standard deviation in the measurement of the spacing of all STO (001) planes in the image is +/-1.5%. In fact the dilation was present in most images and in all the samples observed. But due to noise, it was often difficult to localize, on the one hand, and to precisely measure at different defoci, on the other hand. If defocus never reversed the phenomenon to a contraction, we cannot exclude that it had an influence on the amplitude of the dilation. In another image taken in conditions where all cationic columns were visible in both materials, the dilatation was localized in between the SrO and LaO planes at the interface, i.e. on the $TiO_6$ interfacial octahedra, and amounted to +9 % in the growth direction. However in that more noisy image, the standard deviation in (001) spacing was +/-5%. Although their uncertainty is large, these real-space measurements tend to show that the unit cell patterns are elongated by at least 4% at the interface. The Fourier analyses show the statistical relevance of the phenomenon. Finally, the measurement of this dilatation in HAADF images (Fig. 3) confirms that its nature is not due to TEM artefact.

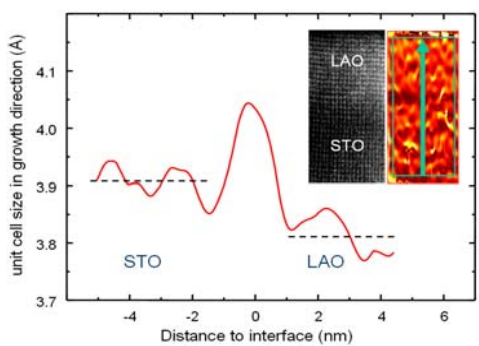

**Fig. 3** (Color online). (001) spacing profile deduced from a STEM-HAADF image of a {110} cross section and its Fourier 002 map (insets). Dilation of interface cells is clearly visible. The horizontal lines give the STO (0.3905 nm), and LAO (0.378 nm) parameters.

## 4 Discussion

We discuss in the following the origin of the image distortion at the interface. We first evaluate the electrostatic charge which the interface possibly carries, and see that it may be at the origin of a part of the distortion. We then discuss an intrinsic distortion of the unit cells at the interface.

In a purely ionic description, electrostatic neutrality at the (001) interface between LAO and $TiO_2$-terminated STO requires the presence of one half extra electron per unit cell area [4-6]. If in practice, other phenomena, implying ion exchange and vacancies [6,15], also come into play, let us stick to such a simple schematic for the sake of clarity. It means that one half of the Ti ions at the interface must have valency 3+. The interface can be considered as made of half-unit cells of $LaTiO_3$; electrostatic equilibrium concerns the reaction: $La^{3+}Ti^{3+}O^{2-}_3 \leftrightarrow La^{3+}Ti^{4+}O^{2-}_3 + e^-$. The conductivity observed in this system is 3-dimensional in character (unpublished magnetoresistance results on the present samples, and [5]), some of the extra electrons should therefore be lacking in the interface plane, as they participate in this conductivity. If so, the number of positively charged $La^{3+}Ti^{4+}O^{2-}_3$ formulae should exceed that of neutral – unionised in the sense of a donor in a semiconductor – $La^{3+}Ti^{3+}O^{2-}_3$. Therefore, there should remain a fixed positive charge at the interface.

Localised charges in the sample may shift images because of Coulomb interaction with the electron beam. If an interface oriented parallel to the beam is charged, the images of the two layers will be shifted in opposite directions, leading to an apparent extension or contraction of the interface patterns. However, these shifts should be proportional to defocus, which provides a means to characterise them [14]. In scanning transmission electron microscopy also, an equivalent effect will occur, as the scanned beam is shifted by the interface potential. We observed changes of the distortion with defocus, but never a reversal as should occur in the case of reversed defocus. We then tend to think that, if interfacial electrostatic charge may play a role in the image distortion, the latter has also, most probably, an intrinsic origin.

Several phenomena could provoke a lattice distortion. One would be that the fixed positive charge at the interface could repulse the positively charged LaO plane. Another explanation, perhaps more seducing, would consider the occupancy of the Ti-3d electron energy levels at the interface. These levels are empty in the case of $Ti^{4+}$, while they contain one electron in the case of $Ti^{3+}$. Solid state chemistry indicates that introducing a single electron in the 3d band of a 3d transition metal in octahedral symmetry will lead, in order to minimize the energy, to a distortion of the octahedron [16,17] by Jahn-Teller effect [18] (Fig. 4).

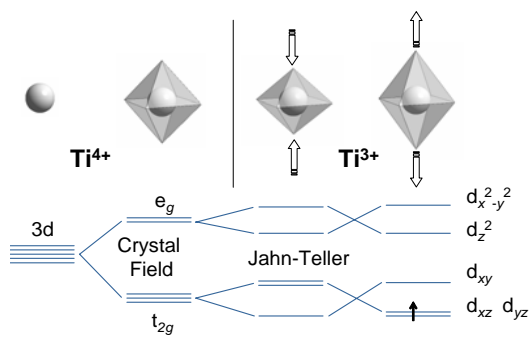

**Fig. 4** Schematic of the correlations between atomic and electronic structures of an $MO_6$ molecule in a crystal, M being a 3d transition metal. Lowering electron energy when introducing an electron in the empty d levels may lead to a contraction of the octahedron by Jahn-Teller effect, or to an elongation, by an effect similar to Jahn-Teller, except that it maintains degeneracy of the base level.

In the present case, the Jahn-Teller effect would promote a contraction of the octahedron. But the dilatation we observe would also minimize electron energy, as shown in Fig. 4, and would give interfacial electrons more degrees of freedom.

## 5 Conclusion

We have grown LAO onto $TiO_2$-terminated STO and found that this association of insulating materials is metallic, confirming the observations of Ohtomo and Hwang [5]. We have then performed atomic-scale imaging of the interface, using $C_S$-corrected TEM on the one hand and HAADF-STEM on the other hand. Our images exhibit a dilation of the unit cells at the interface. This distortion could partly be the result of image shifts associated with interface charge. Its poor sensitivity to defocus in TEM tends to show that it is mainly due to an intrinsic displacement of the atoms. The latter would be due to static charge again or to the presence of extra electrons. EELS experiments and ab initio calculations are in progress to sort these hypotheses.

**Acknowledgements** We would like to thank M. Hÿtch (CNRS, Toulouse) for providing his processing software of HRTEM spatial frequencies, C. Deranlot (UMP CNRS/Thales), for the sputter-deposited contacts, and C. Colliex (CNRS-LPS, Université Paris-Sud, Orsay), M. Bibes (CNRS-IEF, Université Paris-Sud, Orsay), A. Barthélémy and G. Herranz (UMP CNRS/Thales) for fruitful discussions.